%% file: main.tex
\documentclass[letterpaper,twocolumn,10pt]{article}
\usepackage[table]{xcolor}
\usepackage{usenix-2020-09}
\usepackage{setspace}

\usepackage{booktabs}   

\usepackage{fp-frame}
\usepackage{pgffor}
\usepackage{url}
\usepackage{comment}
\usepackage{stmaryrd}
\usepackage{amssymb}
\usepackage{amsmath}
\usepackage{mathtools}
\usepackage{amstext}
\usepackage{mathpartir}
\usepackage{listings}
\usepackage{color}
\usepackage{framed}
\usepackage{tabularx}

\newtheorem{proposition}{Proposition}
\newtheorem{theorem}{Theorem}

\usepackage{graphicx}
\graphicspath{{./visualizations/}}
\usepackage{subcaption}

\input{macros.tex}

\input{ces-macros.tex}

\renewcommand{\paragraph}[1]{\vspace*{1mm}\noindent\textbf{#1}}

\usepackage{enumitem} 
\usepackage{slashed} 
\usepackage{galois}
\usepackage{float}
\usepackage[ruled, noend, linesnumbered]{algorithm2e}

\renewcommand{\arraystretch}{0.75}

\title{Efficient Differentially Private Secure Aggregation for Federated Learning via Hardness of Learning with Errors}

\author{
{\rm Timothy Stevens} \\
University of Vermont
\and
{\rm Christian Skalka}\\
University of Vermont
\and
{\rm Christelle Vincent}\\
University of Vermont
\and
{\rm John Ring}\\
MassMutual
\and
{\rm Samuel Clark}\\
Raytheon
\and
{\rm Joseph Near}\\
University of Vermont
}

\begin{document}

\maketitle

\begin{abstract}
    Federated machine learning leverages edge computing to develop models from network user data, but privacy in federated learning remains a major challenge. Techniques using differential privacy  have been proposed to address this, but bring their own challenges- many require a trusted third party or else add too much noise to produce useful models. Recent advances in \emph{secure aggregation} using multiparty computation eliminate the need for a third party, but are computationally expensive especially at scale. We present a new federated learning protocol that leverages a novel differentially private, malicious secure aggregation protocol based on techniques from Learning With Errors. Our protocol outperforms current state-of-the art techniques, and empirical results show that it scales to a large number of parties, with optimal accuracy for any differentially private federated learning scheme. 
\end{abstract}

\input{intro}

\input{overview}

\input{jfl}

\input{core_protocols}

\input{evaluation_new}

\input{related_work}

\input{conclusion}

\bibliographystyle{plain}
\bibliography{biblio}

\appendix
\input{proof_appendix}

\end{document}

%% file: macros.tex
\newcommand{\fpkeyword}[1]{\text{\sf #1}}

\newcommand{\stack}[1]{\mathord\mid #1\kern.2mm\mathord\mid}

\newcommand{\cod}[1]{\llbracket#1\rrbracket}

\newcommand{\tset}[1]{\{#1\}}
\newcommand{\tnext}{\,;\;}
\newcommand{\drow}[3]{#1:#2\tnext#3}
\newcommand{\urow}[1]{\partial#1}

\newcommand{\capab}[1]{\mathbf{#1}}

\newcommand{\tpre}{\capab{Pre}}
\newcommand{\tabs}{\capab{Abs}}

\newcommand{\system}[2]{\mathcal{S}_{#1}^{#2}}



%% file: intro.tex
\section{Introduction}
\label{sec-intro}

Mobile phones and embedded devices are ubiquitous and allow massive
quantities of data to be collected from users. The recent explosion in
data collection for \emph{deep learning} has led to significant new
capabilities, from image recognition to natural language processing.
But collection of private data from phones and devices remains a major
and growing concern. Even if user data is not directly disclosed,
recent results show that trained models themselves can leak
information about user training data~\cite{shokri2017membership,
  yeom2018privacy}.

Private data for training deep learning models is typically collected
from individual users at a central location, by a party we call the
\emph{server}. But this approach creates a
significant computational burden on data centers, and requires
complete trust in the server.  Many data owners are rightfully
skeptical of this arrangement, and this can impact model accuracy,
since privacy-conscious individuals are likely to withhold some or
even all of their data.

A significant amount of existing research aims to address these issues.
\emph{Federated learning}~\cite{kairouz2019advances} is a family of
decentralized training algorithms for machine learning that allow
individuals to collaboratively train a model without collecting the
training data in a central location. This addresses computational
burden in data centers by shifting training computation to the edge.
However, federated learning does not necessarily protect the privacy
of clients, since the updates received by the server may
reveal information about the client's training data~\cite{shokri2017membership,
  yeom2018privacy}.

Combining \emph{secure aggregation}~\cite{bonawitz2017aggregation,
  bell_paper} with \emph{differential
privacy}~\cite{dwork2014algorithmic, kairouz2021distributed} ensures
end-to-end privacy in federated learning systems. In principle, secure
aggregation allows user updates to be combined without viewing any
single update in isolation. Methods based on differential privacy add
noise to updates to ensure that trained models do not expose
information about training data. However, secure aggregation protocols
are expensive, in terms of both computation and communication. The
state-of-the-art protocol for aggregating large vectors (as in
federated deep learning) is due to Bonawitz et
al.~\cite{bonawitz2017aggregation}. This protocol has a communications
expansion factor of more than 2x when aggregating 500 length-20,000
vectors (i.e.~it doubles the communication required for each
client), and requires several minutes of computation time for the
server.

In this paper we propose a new protocol, called $\ourprot$, that
supports scalable, efficient, and accurate federated learning with
differential privacy, and that does not require a trusted
server. A main technical contribution of our work is a novel
method for \emph{differentially private secure aggregation}. This
method significantly reduces computational overhead as compared to
state-of-the-art-- our protocol reduces
communications expansion factor from 2x to 1.7x for 500 length-20,000
vectors, and reduces computation time for the server to just a few
seconds. The security of this method is is based on the learning with
errors (LWE) problem~\cite{Regev2005}-- intuitively, the noise added for
differential privacy \emph{also} serves as the noise term in LWE.

To obtain computational differential privacy~\cite{mironov2009} $\ourprot$
uses the distributed discrete Gaussian
mechanism~\cite{kairouz2021distributed} and gradient clipping, with
secure aggregation accomplished efficiently via our new method. The
accuracy of our approach is comparable to that achieved by the
\emph{central model} of differential privacy, while providing better
efficiency and thus scalability of previous distributed approaches.
We implement our approach and evaluate it empirically on neural
network architectures for MNIST and CIFAR-10, measuring both accuracy
and scalability of the training procedure. In terms of accuracy, our
results are comparable with central-model approaches for
differentially private deep learning (on MNIST: 95\% accuracy for
$\epsilon \leq 2$; on CIFAR-10: 70\% accuracy for $\epsilon \leq 4$).

\subsection{Contributions} In summary, our contributions are:

\begin{enumerate}
\item A novel malicious-secure aggregation protocol that outperforms
  previous approaches to gradient aggregation with differential privacy.
\item A new end-to-end protocol ($\ourprot$) for privacy-preserving federated learning 
  setting that uses our secure aggregation protocol to provide
  differential privacy even in the presence of malicious clients
  and a malicious server. \item Analytic and empirical results that support our scalability claims, and
  that show our protocol achieves
  nearly the same accuracy as \emph{central-model} approaches for
  differentially private deep learning on practical models for MNIST
  and CIFAR-10. 
\end{enumerate}

%% file: overview.tex
\section{Overview}
\label{sec:overview}
\label{sec:problem}

We study the problem of \emph{distributed differentially private deep learning without a trusted data curator}. Our setting includes a set of \emph{clients} (or data owners), each of whom holds some sensitive data, and a \emph{server} that aggregates gradients generated by clients to obtain a model for the entire federation. The goal is to obtain a differentially private model, without revealing any private data to either the server or other clients.

\begin{table*}[t]
    \begin{center}
    \begin{tabular}{|c||c|c|c|}
        \hline
        \textit{Setting} & \textit{Bonawitz} & \textit{Bell} & \textit{\ourprot} \\ 
        \hline \hline
        Client Communication & $O(k + m)$ & $O(\log k + m)$ & $O(m + k + n)$ \\
        \hline
        Client Computation & $O(k^2 + km)$ & $O(\log^2k + m\log k)$ & $O(mn + k\log k)$ \\
        \hline
        Server Communication & $O(k^2+ km)$ & $O(k\log k + km)$ & $O(mk + n)$\\
        \hline
        Server Computation & $O(mk^2)$ & $O(k\log^2k + km\log k)$ & $O(mk + mn + k\log k)$ \\
        \hline
    \end{tabular}
    \end{center}
    \caption{Communication and computation complexities of \ourprot compared with the state of the art. }
    \label{tab:complex}
\end{table*}

\subsection{Background: General Problem Setting}

\paragraph{Deep learning.}
Deep learning~\cite{goodfellow2016deep} attempts to train a neural network architecture $\mathcal{F}(\theta, \cdot)$ by training its \emph{parameters} (or \emph{weights}) $\theta$ in order to minimize the value of a \emph{loss function} $\mathcal{L}(\theta, \cdot)$ on the training data. Advances in deep learning have lead to significant gains in machine learning capabilities in recent years. Neural networks are typically trained via \emph{gradient descent}: each iteration of training calculates the gradient of the loss on a subset of the training data called a \emph{batch}, and the model parameters are updated based on the negation of the gradient.

Traditional deep learning techniques assume the training data is collected centrally; moreover, recent results suggest that trained models tend to memorize training data, and training examples can later be extracted from the trained model via \emph{membership inference attacks}~\cite{shokri2017membership, yeom2018privacy, carlini2019secret, jayaraman2020revisiting}. When sensitive data is used to train the model, both factors represent significant privacy risks to data owners.

\paragraph{Federated learning.}
Federated learning is a family of techniques for training deep neural networks without collecting the training data centrally. In the simplest form of federated learning (also called \emph{distributed SGD}), each client computes a gradient locally and sends the gradient (instead of the training data) to the server. The server averages the gradients and updates the model. More advanced approaches compute gradients in parallel to reduce communication costs; Kairouz et al.~\cite{kairouz2019advances} provide a survey.

\paragraph{Differentially private deep learning.}
Differential privacy~\cite{dwork2014algorithmic} is a rigorous privacy framework that provides a solution to the problem of privacy attacks on deep learning models. Achieving differential privacy typically involves adding noise to results to ensure privacy. 
Abadi et al.~\cite{dpdl} introduced DP-SGD, an algorithm for training deep neural networks with differential privacy. DP-SGD adds noise to gradients before each model update. Subsequent work has shown that this approach provides strong privacy protection, effectively preventing membership inference attacks~\cite{yeom2018privacy, carlini2019secret, jayaraman2020revisiting}.

DP-SGD works in the \emph{central model} of differential privacy---it requires the training data to be collected centrally (i.e. on a single server). The participant that holds the data and runs the training algorithm is often called the \emph{data curator} or \emph{server}, and in the central model, the server must be trusted. Central-model algorithms offer the best accuracy of known approaches, at the expense of requiring a trusted server.

\paragraph{Federated learning with local differential privacy.}
The classical method to eliminate a trusted server is \emph{local differential privacy}~\cite{dwork2014algorithmic}, in which each client adds noise to their own data before sending it to the server. Local differential privacy algorithms for gradient descent have been proposed, but for deep neural networks, this approach introduces too much noise to train useful models~\cite{bhowmick2018protection}. The major strength of local differential privacy is the threat model: privacy is assured for each client, \emph{even if every other client and the server act maliciously}. The local model of differential privacy has also been relaxed to the \emph{shuffle model}~\cite{cheu2019distributed, erlingsson2019amplification}, which lies between the local and central models but which has seem limited use in distributed machine learning.

\paragraph{Secure aggregation.}
The difference in accuracy between the central and local models raises the question: can cryptography help us obtain the benefits of both, simultaneously? Several \emph{secure aggregation protocols} have been proposed in the context of federated learning to answer this question in the affirmative. These approaches yield the \emph{accuracy of the central model}, but \emph{without a trusted server}.

Secure aggregation protocols allow a group of clients---some of whom may be controlled by a malicious adversary---to compute the \emph{sum} of the clients' privately-held vectors (e.g. gradients, in federated learning), without revealing individual vectors. The state-of-the-art protocol is due to Bonawitz et al.~\cite{bonawitz2017aggregation}. For $k$ clients and length-$m$ vectors, this protocol requires $O(k^2 + mk)$ computation and $O(k + m)$ communication per client, and $O(mk^2)$ computation and $O(m^2 + mk)$  communication for the untrusted server. Bell et al.~\cite{bell_paper} improve these to $O(\log^2k + m \log n)$  computation and $O(\log k + m)$ communication (client) and $O(k \log^2 k + km \log k)$ computation and $O(k \log k + km)$ communication (server). These complexity classifications are summarized in Table~\ref{tab:complex}.

\subsection{Efficient Secure Aggregation in the Differential Privacy Setting}

We present a new protocol for secure aggregation (detailed in Section~\ref{sec-core_protocols}) specifically for the setting of differentially private computations. Our protocol reduces client communications complexity to $O(m + k)$ and server communications complexity to $O(mk)$, where as above we have $k$ parties aggregating vectors of length $m$, and demonstrates excellent concrete performance in our empirical evaluation (Section~\ref{sec:evaluation}).  These analytic results are summarized in Table~\ref{tab:complex} for easy comparison with previous work. 

\paragraph{Threat model.}
Like previous work, we target both the semi-honest setting (in which all clients and the server correctly execute the protocol) and the malicious setting (in which the server and some fraction of the clients may act maliciously). These threat models are standard in the MPC literature~\cite{evans2017pragmatic}, and match the ones targeted by Bonawitz et al.~\cite{bonawitz2017aggregation} and Bell et al.~\cite{bell_paper}. In the semi-honest version, we assume
that the server is honest-but-curious, and that the clients have a
corrupted honest-but-curious subset with an honest majority.  In the
malicious version, we assume that the server is malicious, and that
the clients have a corrupted malicious subset with an honest majority. We present both versions in Section~\ref{sec-core_protocols} (note that the results in Table~\ref{tab:complex} are for semi-honest protocol versions in all cases).

\subsection{Paper Roadmap}

The rest of the paper is organized as follows. In Section
\ref{sec-jfl} we describe the \emph{ideal} but insecure functionality
of our main protocol that assumes a trusted server, along with our
threat model.  The trusted server assumption is removed in Section
\ref{sec-core_protocols} where we present novel techniques for
lightweight malicious-secure aggregation based on LWE. In that Section
we also describe the threat model and state formal security results
for the protocol, and analyze its algorithmic complexity. In Section
\ref{sec:evaluation} we discuss methods and results for two
experiments-one that further evaluates scalability and other
performance parameters, and another that evaluates the accuracy of the
models using our protocol.  We conclude with a summary and remarks on
open related problems in Section \ref{sec:conc}.

%% file: jfl.tex
\newcommand{\fed}{\mathcal{F}}
\newcommand{\aggregator}{\mathcal{A}}

\section{Differentially Private Federated Learning}
\label{sec-jfl}

Abadi et al.~\cite{dpdl} describe a differentially private algorithm
for stochastic gradient descent in the central model of differential
privacy.
The algorithm assumes that the
training data is collected centrally by a trusted curator, and
training takes place on a server controlled by the curator.
For details of the algorithm the reader is referred to~\cite{dpdl} 

The primary challenge in differentially private deep learning is in
bounding the sensitivity of the gradient computation.
Abadi et
al.~\cite{dpdl} use the approach of computing \emph{per-example
  gradients}---one for each example in the minibatch---then
\emph{clipping} each gradient to have $L_2$ norm bounded by the
\emph{clipping parameter} $C$ (line 6). The summation of the clipped
gradients (line 7) has global $L_2$ sensitivity bounded by $C$.

Our privacy analysis of this algorithm uses R\'{e}nyi differential
privacy (RDP)~\cite{mironov2017renyi} (rather than the moments
accountant) for convenience and leverages parallel composition over
the minibatches in each epoch (rather than privacy amplification by
subsampling). Otherwise, it is similar to that of Abadi et al. By the
definition of the Gaussian mechanism for R\'{e}nyi differential
privacy~\cite{mironov2017renyi}, the Gaussian noise added in line 7 is
sufficient to satisfy $\Big(\alpha, \frac{C^2 \alpha}{2 \sigma^2}
\Big)$-RDP. By RDP's sequential composition theorem, training for $E$
epochs satisfies $\Big(\alpha, \frac{E C^2 \alpha}{2 \sigma^2}
\Big)$-RDP. Slightly tighter privacy analyses have been
developed~\cite{dong2019gaussian, bu2020deep, asoodeh2020better} that
also apply to our work. We present the RDP analysis for simplicity,
since our focus is not on improving central-model accuracy.

\subsection{$\ourprot$: Distributed DP SGD}

\begin{algorithm}[t]
  \setstretch{.85}
  \SetAlgorithmName{Protocol}{}{}

  \let\oldnl\nl
  \newcommand{\nonl}{\renewcommand{\nl}{\let\nl\oldnl}}

  \SetKwData{count}{count}
  \SetKwData{Lap}{Lap}
  \SetKwData{noisyCount}{noisyCount}
  \SetKwData{total}{total}
  \SetKwData{pr}{Prob}
  \SetKwInOut{Input}{Input}
  \SetKwInOut{Output}{Output}
  \SetKwFunction{FCU}{ClientUpdate}
  \SetKwFunction{BG}{NoisyBatchGradient}
  \SetKwProg{Fn}{Function}{:}{}

  \nonl \emph{Runs on the \textbf{untrusted server}}\\

  \Input{Set of clients $P$, noise parameter $\sigma$,
    minibatch size $b$, learning rate $\eta$, clipping parameter $C$,
    number of epochs $E$.}
  
  \Output{Noisy model $\theta$.}
  \nonl \textbf{Privacy guarantee:} satisfies $\Big(\alpha, \frac{E C^2 \alpha}{\sigma^2}\Big)$-RDP for $\alpha \geq 1$, assuming honest majority of clients in each batch\\[1mm]

  \vspace*{2mm}

  $\theta \leftarrow \text{random initialization}$\\
  \For{$E$ epochs}{
    \For{each batch of clients $P_b \in P$ of size $b$}{
      $G \leftarrow \BG(P_b, \sigma, C, \theta)$
      $\theta := \theta - \frac{1}{b} \eta G$
      \hfill\textit{update model} \\
    }
  }
  \Return{$\theta$}\\[2mm]

  \caption{$\ourprot$ Protocol}
  \label{alg:distributed_training}
\end{algorithm}

\begin{algorithm}[t]
  \setstretch{.85}
  \SetAlgorithmName{Functionality}{}{}

  \let\oldnl\nl
  \newcommand{\nonl}{\renewcommand{\nl}{\let\nl\oldnl}}

  \SetKwData{count}{count}
  \SetKwData{Lap}{Lap}
  \SetKwData{noisyCount}{noisyCount}
  \SetKwData{total}{total}
  \SetKwData{pr}{Prob}
  \SetKwInOut{Input}{Input}
  \SetKwInOut{Output}{Output}
  \SetKwFunction{FCU}{ClientUpdate}
  \SetKwFunction{BG}{SecureNoisyBatchGradient}
  \SetKwProg{Fn}{Function}{:}{}

  \nonl \hspace*{-2mm} \emph{Runs on a \textbf{trusted third party}}\\[1mm]

  \Input{Batch of clients $P_b$ of size $b$, noise parameter $\sigma$,
    clipping parameter $C$, current model $\theta$.}
  
  \Output{Noisy gradient $\hat{G}$.}

  \nonl \textbf{Privacy guarantee:} satisfies $\Big(\alpha, \frac{C^2 \alpha}{\sigma^2}\Big)$-RDP for $\alpha \geq 1$, assuming honest majority of clients\\[1mm]

  \vspace*{2mm}

  \nonl \hspace*{-3mm} \textbf{Part 1}: each client $p_i \in P_b$ computes a
  noisy gradient and sends it to the functionality $\mathcal{F}$.\\[2mm]
  
  \For{each client $p_i \in P_b$}{
    $g_i \leftarrow \nabla \mathcal{L}(\theta, \mathsf{dataOf}(p_i))$
    \hfill\textit{compute gradient} \\
    $\bar{g}_i \leftarrow g_i / \max(1, \frac{\lVert g_i \rVert_2}{C})$
    \hfill\textit{clip gradient} \\
    $\hat{g}_i \leftarrow \bar{g}_i + \mathcal{N}(0, \frac{\sigma^2}{b} \mathbf{I})$
    \hfill\textit{add noise} \\
    $p_i$ \text{ sends } $\hat{g}_i$ \text{ to } $\mathcal{F}$
  }

  \vspace*{2mm}

  \nonl \hspace*{-3mm} \textbf{Part 2}: $\mathcal{F}$ computes the sum of noisy
  gradients and releases it to the server.\\[2mm]
  
  $\hat{G} \leftarrow \sum_{i = 1}^b \hat{g}_i $
  \hfill\textit{sum individual gradients} \\
  $\mathcal{F}$ \text{ sends } $\hat{G}$ \text{ to the \textbf{untrusted server}}
  
  \caption{Distributed \texttt{NoisyBatchGradient}}
  \label{alg:noisybatchgradient}
\end{algorithm}

We now extend the central-model approach
to the distributed setting. The
following describes a macro-level protocol for realizing
differentially private distributed SGD when a trusted third party is
present. Functionality~\ref{alg:noisybatchgradient}
(\texttt{NoisyBatchGradient}) assumes the existence of a trusted third
party to aggregate the noisy gradients associated with a single batch.
Section~\ref{sec-core_protocols} will describe our MPC protocol that
implements Functionality~\ref{alg:noisybatchgradient} without a
trusted third party.

Together, Protocol~\ref{alg:distributed_training} and
Functionality~\ref{alg:noisybatchgradient} define a differentially
private distributed SGD algorithm suitable for the trusted server
setting.
The distributed computation follows the framework of McMahon et
al.~\cite{bonawitz2017aggregation}, in which each client computes a
gradient locally (Functionality~\ref{alg:noisybatchgradient}, line 2).
To satisfy differential privacy, our adaptation clips each gradient
and adds noise (lines 3-4).

Under the assumption that a trusted third party is available to
compute Functionality~\ref{alg:noisybatchgradient},
Protocol~\ref{alg:distributed_training} satisfies differential
privacy.
Each execution of Functionality~\ref{alg:noisybatchgradient}
calculates a sum of noisy gradients, each with Gaussian noise of scale
$\frac{\sigma}{b}$. The final sum is:
\begin{equation} \label{eq:1}
\hat{G} = \sum_{i=1}^b \hat{g}_i
    = \sum_{i=1}^b \Big(\bar{g}_i + \mathcal{N}(0, \frac{\sigma^2}{b}\textbf{I})\Big)
    = \Big(\sum_{i=1}^b \bar{g}_i\Big) + \mathcal{N}(0, \sigma^2\textbf{I}),
\end{equation}
which is exactly the same as the central model algorithm~\cite{dpdl}.
The last step of the
derivation follows by the sum of Gaussian random variables. Note that
the noise added by each client is \emph{not sufficient} for a
meaningful privacy guarantee (it is only $\frac{1}{b}$ of the noise
required). The privacy guarantee relies on the noise samples being
correctly summed along with the gradients. This is a major difference
between Functionality~\ref{alg:noisybatchgradient} and approaches
based on local differential privacy~\cite{bhowmick2018protection}, in
which \emph{each} client adds sufficient noise for privacy.

The privacy analysis for Functionality~\ref{alg:noisybatchgradient}
and Protocol~\ref{alg:distributed_training} are standard, based on
the conclusion of Equation~\eqref{eq:1}. The $L_2$ sensitivity of
$\Big(\sum_{i=1}^b \bar{g}_i\Big)$ is $C$, since at most one element
of the summation may change, and it may change by at most $C$. By the
definition of the Gaussian mechanism for R\'{e}nyi differential
privacy, the noisy gradient sum satisfies $\Big(\alpha, \frac{C^2
  \alpha}{2 \sigma^2}\Big)$-RDP. The batches are disjoint, so over $E$
epochs of training, each individual in the dataset incurs a total
privacy loss of $\Big(\alpha, \frac{E C^2 \alpha}{2
  \sigma^2}\Big)$-RDP.

\subsection{Security \& Privacy Risks of $\ourprot$}

Protocol~\ref{alg:distributed_training} satisfies differential privacy
when a trusted third party is available to execute
Functionality~\ref{alg:noisybatchgradient}. The server may be
untrusted, since the server only receives differentially private
gradients.

\paragraph{Malicious clients.}
Functionality~\ref{alg:noisybatchgradient} is secure against
semi-honest clients (in part 1), since each client only sees their own
data and the (differentially private) model $\theta$.
However, actively malicious clients may break privacy for \emph{other}
clients. Each client is required to add noise to their own gradient
(line 4); malicious clients may add no noise at all.

If 50\% of the clients add no noise, then the variance of the
noise in the aggregated gradient $\hat{G}$ (line 6) will be
$\frac{\sigma^2}{2}$ instead of $\sigma^2$, yielding $\Big(\alpha,
\frac{E C^2 \alpha}{\sigma^2}\Big)$-RDP (a weaker guarantee than given
above). As the fraction of malicious clients grows, the privacy
guarantee gets weaker. As discussed earlier, we assume an
\textbf{honest majority of clients} and relax our privacy
guarantee to this weaker form.

\paragraph{No trusted third party.}
The larger problem is with the requirement for a trusted third party
to compute Part 2 of Functionality~\ref{alg:noisybatchgradient}. Even
an honest-but-curious server breaks the privacy guarantee for this
part: the server receives each individual gradient separately, and
each one has only a small amount of noise added. This small amount of
noise is insufficient for a meaningful privacy guarantee.
Section~\ref{sec-core_protocols} describes an MPC protocol that
securely implements Functionality~\ref{alg:noisybatchgradient} in the
presence of an actively malicious server and an honest majority of
clients.

\paragraph{Privacy analysis.}
The protocols we describe in Section~\ref{sec-core_protocols} work for
finite field elements, so the floating-point numbers making up noisy
gradients will need to be converted to field elements. Our privacy analysis of
Protocol~\ref{alg:distributed_training} relies on a property of the
sum of Gaussian random variables; as Kairouz et
al.~\cite{kairouz2021distributed} describe, this property does
\emph{not} hold for discrete Gaussians. We amend the privacy analysis
to address this issue in Section~\ref{sec:malic-secure-ourpr}.

%% file: core_protocols.tex
\section{LWE-Based Secure Aggregation}
\label{sec-core_protocols}

In this Section we address the security problem described in the last Section, i.e., that
state-of-the-art federated learning with differential privacy 
requires a trusted third-party server for aggregating gradients. Instead, we propose to use
\emph{secure aggregation}  between the clients of the protocol, eliminating
the need for a trusted third-party server.
This allows us to keep both client inputs and gradients confidential for the
calculation of a differentially private aggregate gradient. Our solution is an secure aggregation protocol that securely realizes Functionality~\ref{alg:noisybatchgradient} as part of Protocol~\ref{alg:distributed_training}.

Our approach is to build a LWE-based masking protocol that
substantially reduces the communication complexity required to add
large vectors. Rather than applying traditional secure
multiparty computation (MPC) protocols to the entire vector, we
generate masks that obscure the secret vectors based on the learning
with errors problem.  The masked vectors are safe to publish to the
central server for aggregation in the clear.  The sum of all vector
masks can be obtained through MPC among the clients in the federation.
Since the individual vector masks cannot be perfectly reconstructed
from the sum of all of the masks, the security of the learning with
errors problem safeguards the encryption of the masked vectors.

Due to the nature of the learning with errors problem, the individual vector masks cannot be perfectly reconstructed with the sum of all the masks. The "errors" remain in the aggregated vector sum, and are sufficient to satisfy $(\epsilon, \delta)$-differential privacy.

\subsection{Background: Learning with Errors}\label{sec:lwebackground}

To reduce the dimension of the vectors that are to be summed using MPC, we use a technique whose security relies on the difficulty of the Learning With Errors (LWE) problems \cite{Regev2005}. These computational problems are usually posed in the following manner: Let $\mathbb{F}_q$ be the finite field of prime size $q$, which is sometimes denoted $GF(q)$, and fix a secret vector $s \in \mathbb{F}_q^n$. An LWE sample is a pair $(a,b)$, where $a \in \mathbb{F}_q^n$ is chosen uniformly at random, and
\begin{equation*}
b = a \cdot s + e \in \mathbb{F}_q,
\end{equation*}
where $a \cdot s$ denotes the usual dot product, and $e$ is a so-called ``error," chosen from a suitable error distribution $\chi$ on $\mathbb{F}_q$. Then the LWE (search) problem consists of retrieving the secret $s$ given a polynomial number of LWE samples $(a,b)$.

For our purposes we will also need the hardness of the LWE decision problem, which is the problem of distinguishing a set of pairs $(a,b)$ with each pair chosen uniformly at random from $\mathbb{F}_q^n \times \mathbb{F}_q$ from a set of pairs that are LWE samples. In \cite{Regev2005}, Regev shows that when $q$ is a prime of size polynomial in $n$ and for $\chi$ any error distribution on $\mathbb{F}_q$, the LWE decision problem is at least as hard as the LWE search problem. Since the reduction from the LWE decision to the LWE search problem is trivial, in those cases the two problems are equivalent.

\subsection{Background: Multiparty Computation}
Secure Multiparty Computation, abbreviated MPC, refers to distributed protocols
where independent data owners use cryptography to compute a shared function output
without revealing their private inputs to each other or a third party \cite{evans2017pragmatic}.
In our setting, the ideal functionality computed by these clients is gradient
aggregation, which as discussed in Section~\ref{sec-jfl} is differentially private with regard to
user inputs. 
Thus MPC serves to replace a trusted third party in secure function evaluation.

Security properties of Secure Aggregation protocols are categorized based on assumptions about the power of an adversary.
\textit{Semi-Honest} adversaries perform the protocol as intended, while attempting to gain information about the private inputs of the protocol.
\textit{Malicious} adversaries may exhibit arbitrary behaviors to affect the security, correctness, or fairness of an MPC protocol.
Furthermore, MPC protocols must assume that some proportion of the involved clients are honest.
\ourprot assumes an honest majority against a malicious adversary.
For a group of size $k$, we assume that $\frac{k}{2} + 1$ clients are honest, and make no assumption about the behavior of the rest.

\ourprot requires the realization of secure vector aggregation in order to add the secret keys each participant uses to mask their larger dimension vectors. Several secure vector aggregation protocols already exist, especially for smaller sized vectors~\cite{bell_paper, bonawitz2017aggregation, shamir1979share}. For the sake of consistent security and complexity analysis, we implement a secure vector aggregation protocol using Shamir secret sharing:

A $(t, k)$ threshold secret sharing scheme will break a secret value into $k$ shares, and require at least $t$ shares to recover the secret. Our secure vector aggregation protocol additionally requires that the scheme have an additive homomorphic property. That is to say if \texttt{[a]} and \texttt{[b]} are secret shares of values $a$ and $b$, and $c$ is a constant. 
        Using \texttt{[a]}, \texttt{[b]}, and $c$, a party must be able to calculate \texttt{[a + b]}, \texttt{[ac]}, and \texttt{[a + c]} without communication among the other clients.

Shamir secret sharing~\cite{shamir1979share} provides this property. Packed Shamir sharing~\cite{franklinyung} is an extension of the original scheme that improves communication efficiency by allowing up to $k - t - 1$ values to be included in one set of shares. This packed variant is used in \ourprot.

\subsection{LWE-Based Masking of Input Vectors}

We now describe our novel masking protocol, which allows us to reduce client communication. A high-level summary of the protocol is the following:
\begin{enumerate}[leftmargin=5mm, itemsep=0pt]
\item Each client generates a one-time-pad that is the same size as their gradient, masks their gradient, and sends the encrypted gradient to the server.
\item Clients add their masks together using MPC and send the aggregate mask to the server.
\end{enumerate}
Through this protocol the server can recover the true sum of the gradients by adding the masked gradients and subtracting the aggregate mask. Moreover, the aggregate mask reveals nothing about any individual gradients or their masks.

We begin by assuming that all clients to the communication share a public set of $m$ vectors chosen uniformly at random from $\mathbb{F}_q^n$, and we arrange these vectors as the rows of an $m \times n$ matrix $A \in \mathbb{F}_q^{m\times n}$. Then each client generates a secret vector $s \in \mathbb{F}_q^n,$ with each entry of the vector drawn from the distribution $\chi$, and an error vector $e \in \mathbb{F}_q^m$, with each entry of the vector also drawn from the same distribution $\chi$, and computes the vector
\begin{equation*}
b = As + e \in \mathbb{F}_q^m.
\end{equation*}

We can then think of the pair $(A,b)$ as a set of $m$ LWE samples, where each row of $A$ constitutes the first entry of a sample as described in Section~\ref{sec:lwebackground}, and each entry of $b$ constitutes the second entry of the sample. The hardness of the LWE decision problem tells us that the vector $b$ is indistinguishable from a vector whose entries are chosen uniformly at random from $\mathbb{F}_q$, so $b$ can serve as a one-time pad to encrypt the vector $v \in \mathbb{F}_q^m$:
\begin{equation*}
 h = v + b,
\end{equation*}
where here $h$ is used to denote the encrypted $v$. Note that according to Regev \cite{Regev2005}, there is no loss in security in having all clients share the same matrix $A$ to perform this part of the protocol.

Now suppose that $h_i$, $v_i$, $b_i$, $s_i$, and $e_i$ are the $h$, $v$, $b$, $s$, and $e$ vectors of client $i$. Additionally, suppose $h_{sum}$, $v_{sum}$, $b_{sum}$, $s_{sum}$ and $e_{sum}$ are the sum of all $b_i$, $s_i$, and $e_i$ for clients $0, \ldots, k-1$ where $k$ is the number of clients.

By the definition of one-time pads, each client can send $h_i$ to the server without revealing anything about $v_i$. The server can obtain $h_{sum}$ through simple vector addition. By the definition of each $h_i$, we further know that:
\[
    h_{sum} = v_{sum} + b_{sum},
\]
and by the definition of each $b_i$ and the distributive property, we obtain:
\[
    h_{sum} = v_{sum} + As_{sum} + e_{sum},
\]
where $As_{sum}$ denotes the usual matrix-vector multiplication. To obtain $s_{sum}$ we assume the federation has access to a secure aggregation protocol that realizes functionality $\texttt{Sagg}(x_0, \dots x_k, t)$. $\texttt{Sagg}$ returns the sum of vectors $x_0, \dots, x_k$, while not revealing any information about any inputs to any subset of parties of size smaller than $t$. Because they utilize $\texttt{Sagg}$, this reveals nothing about their individual $s_i$ values. In the case of dropouts, $\texttt{Sagg}$ also returns the subset of parties that participated in the aggregation. Using $s_{sum}$, the server can compute the following value:

\[
     v_{sum} + e_{sum}
\]
Of course, the clients do not share their individual error vector $e_i$ values because this would invalidate the LWE assumption that ensures $b_i$ is a one-time pad. Therefore, we realize the ideal functionality of calculating $v_{sum}$ by returning a noisy answer. Fortunately, each entry in $e_{sum}$ is the sum of at most $k$ discretized Gaussians.  Therefore we can use the noise added by $e_{sum}$ to satisfy $(\epsilon, \delta)$-DP.

\begin{algorithm} 
  \setstretch{.8}
  \newcommand{\nonl}{\renewcommand{\nl}{\let\nl\oldnl}}
    \SetAlgorithmName{Protocol}{}{}
    \SetKwInOut{Input}{Input}
    \SetKwInOut{Output}{Output}

    \Input{Set $U$  of $k$ clients, each client $i$ has a vector $v_i \in \mathbb{F}_q^m$, the secret length $n$, an error distribution $\chi$, and a common matrix $A \in \mathbb{F}_q^{m \times n}$.}
    \Output{The sum of all vectors $v_0 \dots v_{k-1}$, $V$ }
    {\nonl \textbf{Round 1:} Each client $i$:}
    \begin{enumerate}[itemsep=-1mm]
        \item generates a vector $s_i \in \mathbb{F}_q^{n}$, with each entry drawn at random from $\chi$, using a secret seed.
        \item generates $e_i \in \mathbb{F}^{m}$ with each entry drawn at random from $\chi$.
        \item $b_i \leftarrow As_i + e_i$
        \item $h_i \leftarrow  v_i + b_i$
        \item sends $h_i$ to the server.
    \end{enumerate}
    {\nonl \textbf{Round 2:} The server:}
    \begin{enumerate}[itemsep=-1mm]
        \item receives $h_i$ from each non-dropped out client
        \item the server sends each party the set of clients who sent an $h$. Call this set $U_1$.
    \end{enumerate}
    {\nonl \textbf{Round 3:} Each client $i$:}
    \begin{enumerate}[itemsep=-1mm]
        \item Obtains $s \leftarrow \sum_{i \in U_1} s_i$. Using $\texttt{Sagg}(\{s_i | i \in U_1\}, t)$ and $U_2$, the set of clients that participated in $\texttt{Sagg}$.
        \item sends $s,\ U_2$ to server.
    \end{enumerate}
    {\nonl \textbf{Round 4:} The server:}
    \begin{enumerate}[itemsep=-1mm]
        \item $H \leftarrow \sum_{i \in U_2} h_i$ 
        \item $V \leftarrow H - As$
    \end{enumerate}
    \caption{Masking Aggregation}
    \label{prot:mask}
\end{algorithm}

Protocol \ref{prot:mask} reduces the client communication complexity from $O(\log(q)mk)$ to $O(\log(q)(m + n + k))$ by requiring clients to securely aggregate only a small vector of size $n$. The addition of $n$ and $k$ can be attributed to the possible use of packed secret sharing. Each client shares their length-$m$ vector once with the server, and then uses a packed secret sharing scheme on their length-$n$ vector. The total number of shares required in the packed scheme is $O(n + k)$

\subsection{Vector Aggregation}
To add the secret vectors $s_0 \dots s_{k-1}$, we can use any secure aggregation protocol. In our use cases, each $s_i$ is typically of small dimension $(m \le 800)$, so we use a packed Shamir secret sharing protocol outlined in Protocol~\ref{prot:vec}.

\begin{algorithm}
  \setstretch{.8}
  \newcommand{\nonl}{\renewcommand{\nl}{\let\nl\oldnl}}

    \SetAlgorithmName{Protocol}{}{}
    \SetKwInOut{Input}{Input}
    \SetKwInOut{Output}{Output}
    \Input{$k$ vectors $v_i \in \mathbb{F}_q^n$, one from each client $P_i$,
           a secret sharing threshold $t$, a packing threshold $p < k - t - 1$.}
    \Output{vector sum $V \in \mathbb{F}_q^n$}
    \nonl \textbf{Round 1:} Each client $j$:
    \begin{enumerate}[itemsep=-.5mm]
        \item partitions $v_j$ into a set of length-$p$ vectors $R_j$
        \item Generates a set of $(t-p+1, t+1, p, k)$-packed secret sharing called $S_j$ with one sharing for each vector in $R_j$.
        \item Distributes the shares of each sharing in $S_j$ to clients $P_0 \ldots P_{k-1}$. For a given sharing $s$ in $S_j$, $P_i$ receives $s_{ij}$.
    \end{enumerate}
    \textbf{Round 2:} Each client $j$:
    \begin{enumerate}[itemsep=-.5mm]
        \item Receives shares $s_{j0}, \ldots s_{jk-1}$ from $P_0, \ldots P_{k-1}$ for all sharings in $S$.
        \item $sum_j \leftarrow \sum s_{j0}, \ldots, s_{jk-1}$ for each sharing in $S$.
        \item Broadcasts each $sum_j$ to every client.
    \end{enumerate}
    \textbf{Round 3:} Each client $j$:
    \begin{enumerate}[itemsep=-.5mm]
        \item Receives $sum_0$ \dots $sum_{k-1}$ for each sharing in $S$.
        \item Runs \texttt{reconstruct} on each element $sum_0$ \dots $sum_{k-1}$ to obtain a set of length-$p$ vectors $R_{sum}$.
        \item if (2) fails, broadcast \texttt{ABORT}.
        \item concatenate the vectors in $R_{sum}$ to obtain $V$.
    \end{enumerate}
    \caption{Secure Vector Addition}
    \label{prot:vec}
\end{algorithm}

Protocol~\ref{prot:vec} is secure against semi-honest adversaries based on the security packed secret sharing. A malicious adversary could broadcast an incorrect sum in Round 2 of the protocol, and the final result would be calculated incorrectly by the other clients. 
Traditionally, the \texttt{reconstruct} function has no ability to catch this kind of cheating; in many cases \emph{all} of the shares are needed to reach the threshold during reconstruction, so corruption of a single one will change the result.

\subsection{Malicious-Secure Vector Aggregation}

We now extend Protocol~\ref{prot:vec} to be secure against malicious clients by applying a variation of Benaloh's verifiable secret scheme~\cite{benaloh86}. The key insight behind this modification comes from the observation that in our protocol each client receives $k$ shares  from the other clients in Round 3, but only $t$ shares are actually required for reconstruction. Our modified reconstruction procedure uses the remaining shares to catch cheating clients.

\begin{algorithm}
  \setstretch{.85}
    \SetKwInOut{Input}{Input}
    \SetKwInOut{Output}{Output}
    \Input{Let $[a]$ be a $(t, k)$-Shamir sharing of secret $a$. Assume one client has access to at least $t + 1$ shares of $[a]$.}
    \Output{$a$ or \texttt{ABORT}}
    
    $ A \subset B \subseteq [a]$ where $|A| = t$ and $|B| = t + 1$. \\
    $a' \rightarrow \texttt{reconstruct($A$)} $\\ 
    $b  \rightarrow \texttt{reconstruct($B$)} $\\
    \uIf{$a' = b$}{
        \Return{a'}}
    \uElse{
        \Return{\texttt{ABORT}}
    }
    \caption{Shamir Reconstruction with Verification}
    \label{alg:recon}
\end{algorithm}

We propose the following reconstruction method for verifying that clients have behaved honestly.
Requiring that each client has at least $t + 1$ shares, we have each honest client take two subsets of the shares, one of size $t$ and one of size $t + 1$. 
The clients perform the traditional reconstruction technique on both subsets.
If the values returned by both reconstructions are equivalent, they accept the result as correct. 
Otherwise, they abort.
The modified reconstruction procedure appears in Algorithm~\ref{alg:recon}.
Replacing the call to \texttt{reconstruct} in Protocol~\ref{prot:vec} with a call to this modified reconstruction procedure yields a malicious-secure protocol.

Note that Algorithm~\ref{alg:recon} does not require communication with other clients.
General-purpose malicious-secure protocols based on the same principle require interaction between the clients to check for cheating (e.g., the protocol of Chida et al.~\cite{chida2018fast}) because they use the ``extra'' shares to perform multiplication. Since our application does not require multiplication, we can use these shares to catch cheating instead.

Algorithm~\ref{alg:recon}  can be extended to the packed Shamir variant by requiring that each client has access to $t + k + 1$ shares. 
The number of shares to which access is required must be increased because the reconstruction threshold is increased in the packed variant.
Protocol~\ref{prot:vec} and Algorithm~\ref{alg:recon} realize the ideal functionality $\texttt{Sagg}$ in the malicious adversary threat model.

\subsection{Security Analysis}

Here we analyze the security of Protocol~\ref{prot:mask}, which we will denote as $\pi$.

Suppose the ideal functionality of noisy vector addition as $F$, an adversary $A$. Let $v_i$ and $x_i$ be input and view of client $i$ respectively. Let $x_s$ be the view of the server. $n$ is the LWE security parameter. Suppose a maliciously secure aggregation protocol $\texttt{Sagg}(X, t)$.  Let $V$ be the output of $\pi$.

Let $U$ be the set of clients, and $C \subset U \cup  \{S\}$ be the set of corrupt parties.

In the malicious model, we consider dropping out an adversarial behavior without loss of generality.

Suppose the simulator has access to an oracle $\texttt{IDEAL}(t, v_u)_{u \in U \setminus C}$ where:

\[ \texttt{IDEAL}(t, v_u)_{u \in U \setminus C} = \begin{cases}
    \Sigma_{u \in U \setminus C} v_u & | U \setminus C | >  t \\
    \bot & otherwise
\end{cases}
\]

Let $\texttt{REAL}_{\pi,C}^U = \{x_i | i \in C\},V$.

\begin{theorem}
     There exists a PPT simulator \texttt{SIM} such that for all $t$, $U$, $C$

\[ \texttt{REAL}_{\pi, C}^U(n, t; v_{U \setminus C}) \equiv \texttt{SIM}_C^{U, \texttt{IDEAL}(t, v_u)}(n, t; x_C) \]
 
\end{theorem}

The proof full proof of this theorem can be found in Appendix~\ref{app:proof}.

\subsubsection{LWE parameters}\label{sec:LWEparameters}

The security of an LWE instance is parameterized by the tuple $(n,q, \beta)$ where $n$ is the width of the matrix $A$ (or equivalently the dimension of the secret $s$), $q$ is the field size, and $\beta$ is such that $\beta q$ is the width of the error distribution $\chi$ (so that the standard deviation is $\sigma = \frac{\beta q}{\sqrt{2\pi}}$; this quantity is denoted $\alpha$ in the LWE literature, but we choose $\beta$ here so as to not conflict with the notation for R\'{e}nyi divergence). We used the LWE estimator \cite{estimator} to calculate the security of each parameter tuple. Table~\ref{tab:mask_time}  displays a series of LWE parameters for different potential aggregation scenarios, each with at least 128 bits of security.

The different parameter settings are driven by different sizes of $q$, which would enable more precision in the aggregate values. A larger field size also allows more clients to be involved in the aggregation. Field sizes picked here may also utilize fast Fourier transform secret sharing. For this reason we consider $q$ fixed by the application of the protocol. Since we also use a fixed valued of $\beta = \frac{3.2}{q}$, the security offered by the LWE problem depends on the variable $n$ (the length of the secret $s$), which we call the \emph{security parameter}.

\subsection{Encoding and Decoding Gradients}

In order to manipulate gradients with MPC, we require that they can be encoded as a vector of finite field elements. 
First we flatten the tensors that compose each gradient into a vector of floating point numbers. 
The aggregation operation of gradients is element wise. 
Therefore, we simplify the encoding problem to encoding a floating point number as a finite field element. 
Gradient elements are clipped, and encoded as fixed point numbers. We chose 16 bit numbers with 4 digits of precision after the decimal. This precision was sufficient for model conversion on the MNIST and CFAR-10 problems. 

The integers are converted to unsigned integers using an offset, and the unsigned integer result can be encoded into any field larger than $2^{16}$. The fields used in our experiment are outlined in Table~\ref{tab:mask_time}.

\subsection{Malicious Secure \ourprot}
\label{sec:malic-secure-ourpr}

We now have all the MPC operations necessary to implement our ideal functionality from Protocol~\ref{alg:distributed_training} as a secure multiparty computation.
Protocol~\ref{alg:malicious_batch_gradient} securely implements Functionality~\ref{alg:noisybatchgradient}, and can replace it directly to implement $\ourprot$.
This version of \texttt{NoisyBatchGradient} computes the gradient and adds noise to it in the same way as the ideal functionality, but invokes Protocol~\ref{prot:vec} to sum the vectors. This requires encoding each noisy gradient as a vector of field elements, as described in the last section.

\begin{algorithm}[t]
  \setstretch{.85}
  \SetAlgorithmName{Protocol}{}{}

  \let\oldnl\nl
  \newcommand{\nonl}{\renewcommand{\nl}{\let\nl\oldnl}}

  \SetKwInOut{Input}{Input}
  \SetKwInOut{Output}{Output}
  \SetKwFunction{FCU}{ClientUpdate}
  \SetKwFunction{BG}{SecureBatchGradient}
  \SetKwFunction{EG}{EncodeGradient}
  \SetKwFunction{DG}{DecodeGradient}
  \SetKwFunction{SVA}{SecureVectorAddition}
  \SetKwProg{Fn}{Function}{:}{}
  
  \Input{Batch of clients $P_k$ of size $k$, noise parameter $\sigma$,
    clipping parameter $C$, current model $\theta$.}
  
  \Output{Noisy gradient $\hat{G}$.}

    \nonl \textbf{Privacy guarantee:} satisfies $\Big(\alpha, \frac{C^2 \alpha}{\sigma^2}\Big)$-RDP for $\alpha \geq 1$, assuming honest majority of clients\\[1mm]

  \vspace*{2mm}

  \For{each client $p_i \in P_k$}{
    $g_i \leftarrow \nabla \mathcal{L}(\theta, \mathsf{dataOf}(p_i))$
    \hfill\textit{compute gradient} \\
    $\bar{g}_i \leftarrow g_i / \max(1, \frac{\lVert g_i \rVert_2}{C})$
    \hfill\textit{clip gradient} \\
    $\hat{g}_i \leftarrow \bar{g}_i + \mathcal{N}(0, \frac{\sigma^2}{k} \mathbf{I})$
    \hfill\textit{add noise} \\
    $v_i \leftarrow  \EG(\hat{g_i})$
    \hfill\textit{encode gradient} \\
  }

  \vspace*{2mm}

  \nonl Client $p_i \in P_k$ provides $v_i$ as input to Secure Vector
  Addition (Protocol~\ref{prot:vec}). Together, the clients compute
  $\hat{G} = \sum_{i=1}^k \hat{g}_i$. $\hat{G}$ is released to the
  \textbf{untrusted server}.

  \caption{Malicious-Secure \texttt{NoisyBatchGradient}}
  \label{alg:malicious_batch_gradient}
\end{algorithm}

\paragraph{Privacy analysis.}
The privacy analysis of Protocol~\ref{alg:distributed_training} relies
on the fact that the sum of Gaussian random variables is itself a
Gaussian random variable. However, as Kairouz et
al.~\cite{kairouz2021distributed} point out, this property does not
hold for \emph{discrete} Gaussians---and since \texttt{EncodeGradient}
uses a fixed-point representation for noisy gradients, we cannot rely
on the summation property.
Instead, our privacy analysis proceeds based on Proposition 14 of
Kairouz et al.~\cite{kairouz2021distributed}:
\begin{proposition}[from Kairouz et al.~\cite{kairouz2021distributed}]
  Let $\sigma \geq \frac{1}{2}$. Let $X_{i,j} \leftarrow
  \mathcal{N}_\mathbb{Z}(0, \sigma^2)$ independently for each $i$ and
  $j$. Let $X_i = (X_{i, 1}, \dots, X_{i,d}) \in \mathbb{Z}^d$. Let
  $Z_n = \sum_{i=1}^n X_i \in \mathbb{Z}^d$. Then, for all $\Delta \in
  \mathbb{Z}^d$ and all $\alpha \in [1, \infty)$,
  \[ D_\alpha(Z_n || Z_n + \Delta) \leq \frac{\alpha \lVert \Delta
      \rVert_2^2}{2n\sigma^2} + \tau d\]
  where $\tau := 10 \cdot \sum_{k=1}^{n-1}
  e^{-2\pi^2\sigma^2\frac{k}{k+1}}$.
  \label{prop:discrete}
\end{proposition}

Proposition~\ref{prop:discrete} provides a bound on R\'{e}nyi
divergence, $D_{\alpha}$, for noise generated as the sum of discrete Gaussians, which
directly implies R\'{e}nyi differential privacy.
In our setting, Proposition~\ref{prop:discrete} yields almost
identical results to the privacy analysis of
Protocol~\ref{alg:distributed_training} (which assumes continuous
Gaussians). Note that the first term of the bound from
Proposition~\ref{prop:discrete} is identical to the bound given in our
earlier privacy analysis, when $n$ is equal to the batch size $b$ and
$\lVert \Delta \rVert_2^2$ is equal to $C^2$ (where $C$ is the
$L_2$ clipping parameter). 

As the fixed-point representation of noisy gradients becomes more
precise, the second term of the bound ($\tau d$) becomes extremely
small. The \texttt{EncodeGradient} function uses 4 places of precision
past the decimal point, meaning that the effective values of
$\sigma^2$ and $\lVert \Delta \rVert_2^2$ are 10,000 times their
``original'' values. Each additional place of precision adds another
factor of 10 to both values. This has the effect of reducing the value
of $\tau$ to extremely close to zero.

We have implemented both the original analysis (which incorrectly
assumes continuous Gaussians) and Proposition~\ref{prop:discrete}. The
results reported in Section~\ref{sec:evaluation} use
Proposition~\ref{prop:discrete}, but the two methods yield values of
$\epsilon$ so close together that the resulting graphs are
indistinguishable.

       \subsection{Algorithmic Complexity}

Client computation is comprised of three tasks. Generating a random vector $s$ of length $n$, generating a random vector $e$ of length $m$, multiplying $s$ by $m \times n$ matrix $A$, and generating secret shares for $s$. Random vector generation is an $O(m + n)$ operation where $n$ is length of secret vector $s$ and $m$ is the length of $e$. This can be reduced to $O(m)$ because $m$ will be larger than $n$ in any practical use of \ourprot. Matrix multiplication by a vector is an $O(mn)$ operation where $m$ is the vector size; each matrix element is considered exactly once. Finally, secret share generation is done using the packed FFT method \cite{dahl_2017}, and therefore has a complexity of $O(k\log(k))$ where $k$ is the number of clients. In sum, this gives us a runtime of $O(mn + k\log(k))$ for client computation with respect to our vector size $m$ and $s$ length $n$. 

In order to assume the difficulty of the LWE decision problem, we require that $q$ be polynomial in $n$. Though the field size does affect the precision of the values to be aggregated and the possible number of parties to the aggregation scheme, it is customary to think of $q$ as a constant, and therefore $n$ is constant here too in our complexity analysis. However, in practice it is possible to choose $n$ quite small relative to $q$.

Server complexity consists of adding $k$ masked vectors, reconstructing the packed secret sharing, and multiplying an $m \times n$ matrix by a length $n$ vector. The vector addition and matrix multiplication have complexity $O(mk + m\log(k))$. Reconstructing the packed secret shares takes time $O(k\log(k))$ in the semi-honest case with no dropouts using the Fast Fourier Transform method. In the case of malicious security and dropouts, we use Lagrange interpolation to obtain a runtime of $O(k^2)$. The number of dropouts does not affect runtime complexity as long as there are more than 0 dropouts. In total, the server runtime complexity is $O(mk + mn + k\log(k))$ in the no dropout scenario, and $O(mk + mn + k^2)$ in case of dropouts or malicious adversaries.

%% file: evaluation_new.tex
\section{Evaluation}
\label{sec:evaluation}

Our empirical evaluation aims to answer two research questions:

\begin{itemize}
\item \textbf{RQ1}: How does the concrete performance of \ourprot compare to state-of-the-art secure aggregation?
\item \textbf{RQ2}: Is \ourprot capable of training accurate models?
\end{itemize}

\noindent We conduct two experiments to answer both questions in the
affirmative. We first describe our experiment setup and the datasets
used in our evaluation. Section~\ref{eval:mpc} describes our
scalability experiment; the results show that \ourprot scales to
realistic batch sizes, and that model updates take only seconds.
Section~\ref{eval:ml} describes our accuracy experiment; the results
demonstrate that \ourprot trains models with comparable accuracy to
central-model differentially private training algorithms.

\begin{figure*}[t]
    \begin{subfigure}{.5\textwidth}
        \includegraphics[width=.9\linewidth]{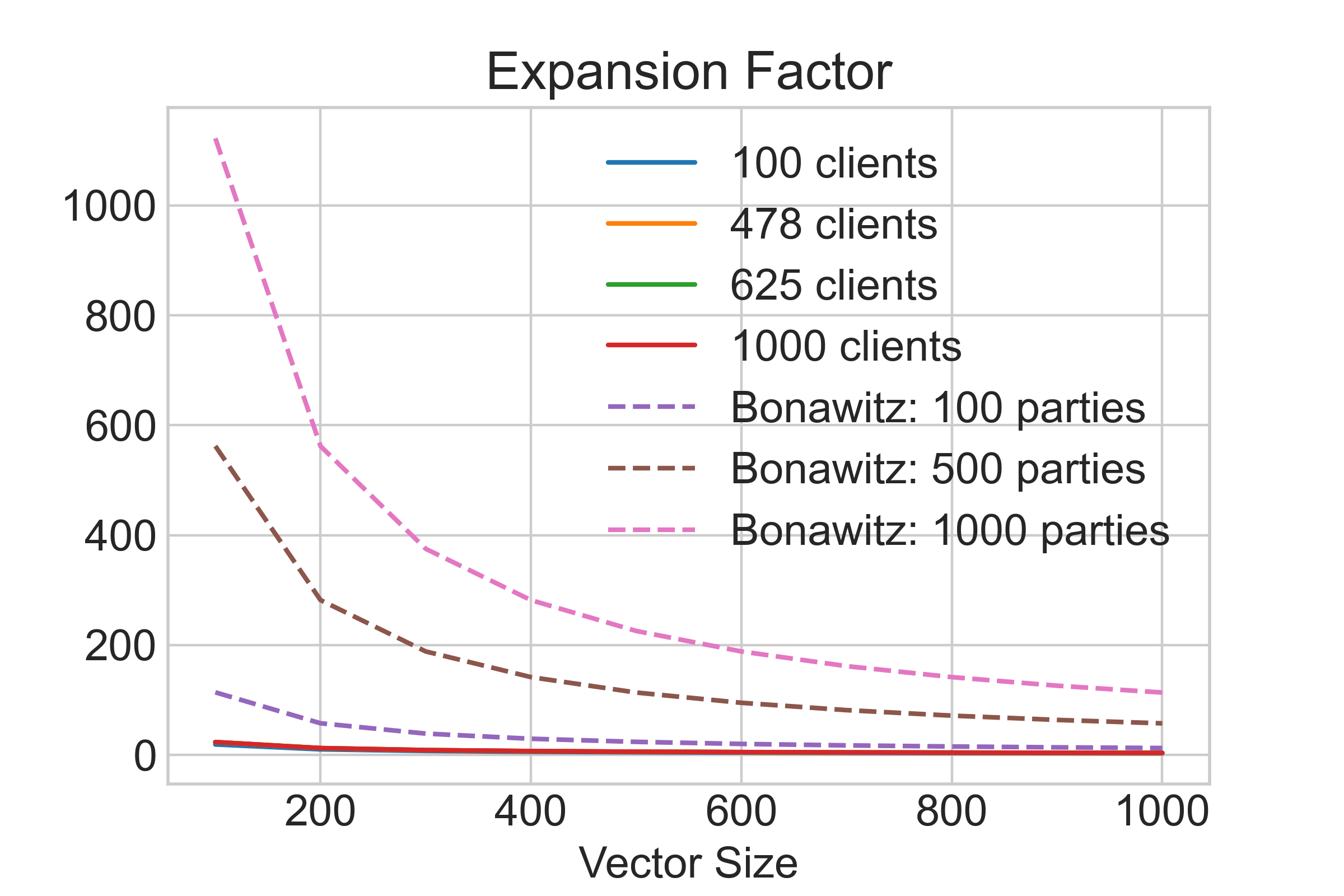}
    \end{subfigure}
    \begin{subfigure}{.5\textwidth}
        \includegraphics[width=.9\linewidth]{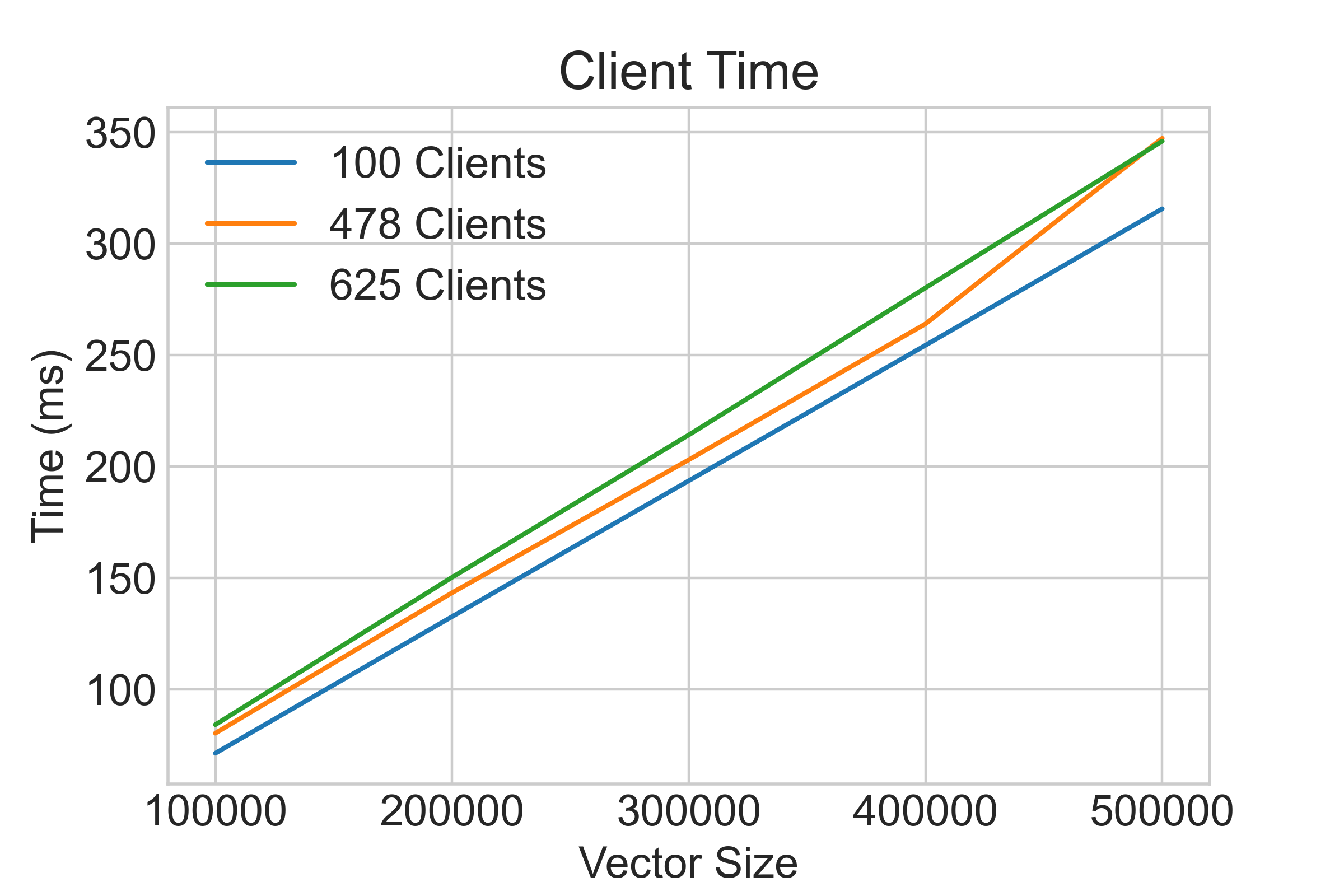}
    \end{subfigure}
    \caption{
    The left figure displays the expansion factor for using our protocol with various vector sizes and numbers of clients, comparing our approach (solid lines) against the secure aggregation protocol of Bonawitz et al.~\cite{bonawitz2017aggregation} (dashed lines).
    The right figure includes cost of a single client's computation. The client timing results are identical regardless of the dropout so only the dropout situation is plotted. 
}
    \label{exp:fig}
\end{figure*}

\paragraph{Experiment setup.}
Our experiments take place in two phases. 
First, the model is trained in a single process with privacy preserving noise added to each gradient.
As model training occurs, gradients for each training sample are written to file.
The second phase involves the MPC simulation.
Clients read their noisy gradients from file, and aggregate them using \ourprot.

This experimental setup is necessary for the implementation of local experiments with batch sizes of $128$. 
Reading each gradient from file sidesteps the need for each client to have their own TensorFlow instance, substantially reducing our memory consumption footprint.

Running these two separate experiments ensures that the MPC results reflect the performance of \ourprot without considering the overhead of training $128$ separate neural networks in parallel.

The memory consumption issue described here is created by simulating many clients on the same machine.  
In a true federated learning instance, each client would have their own independent resources, and therefore would not run into this same issue.


\subsection{Experiment 1: Masking Scalability}
\label{eval:mpc}
This section strives to answer \textbf{RQ1}. 
We implemented the masking protocol in single threaded python and evaluated various federation configurations.
Experiments were run on an AWS z1d2xlarge instance with a 4.0Ghz Intel Xeon processor and 64 Gb of RAM \cite{aws}.
Concrete timing and expansion results for protocol computation are included in Figures~\ref{exp:fig},~\ref{fig:server:time}, and Table~\ref{tab:mask_time}. 
We assume semi-honest behavior from the adversary and consider the scenario with no dropouts as well as a $25\%$ dropout rate. 
In all experiments, $\beta$  is assumed to be $3.2/q$.
We assume a single aggregation server, and we assume that clients broadcast the sum of shares to the server rather than performing Shamir reconstruction themselves.

\subsubsection{Experimental Performance}\label{sec:exp1performance}
\begin{figure*}[t]
    \begin{subfigure}{.5\textwidth}
        \includegraphics[width=.9\linewidth]{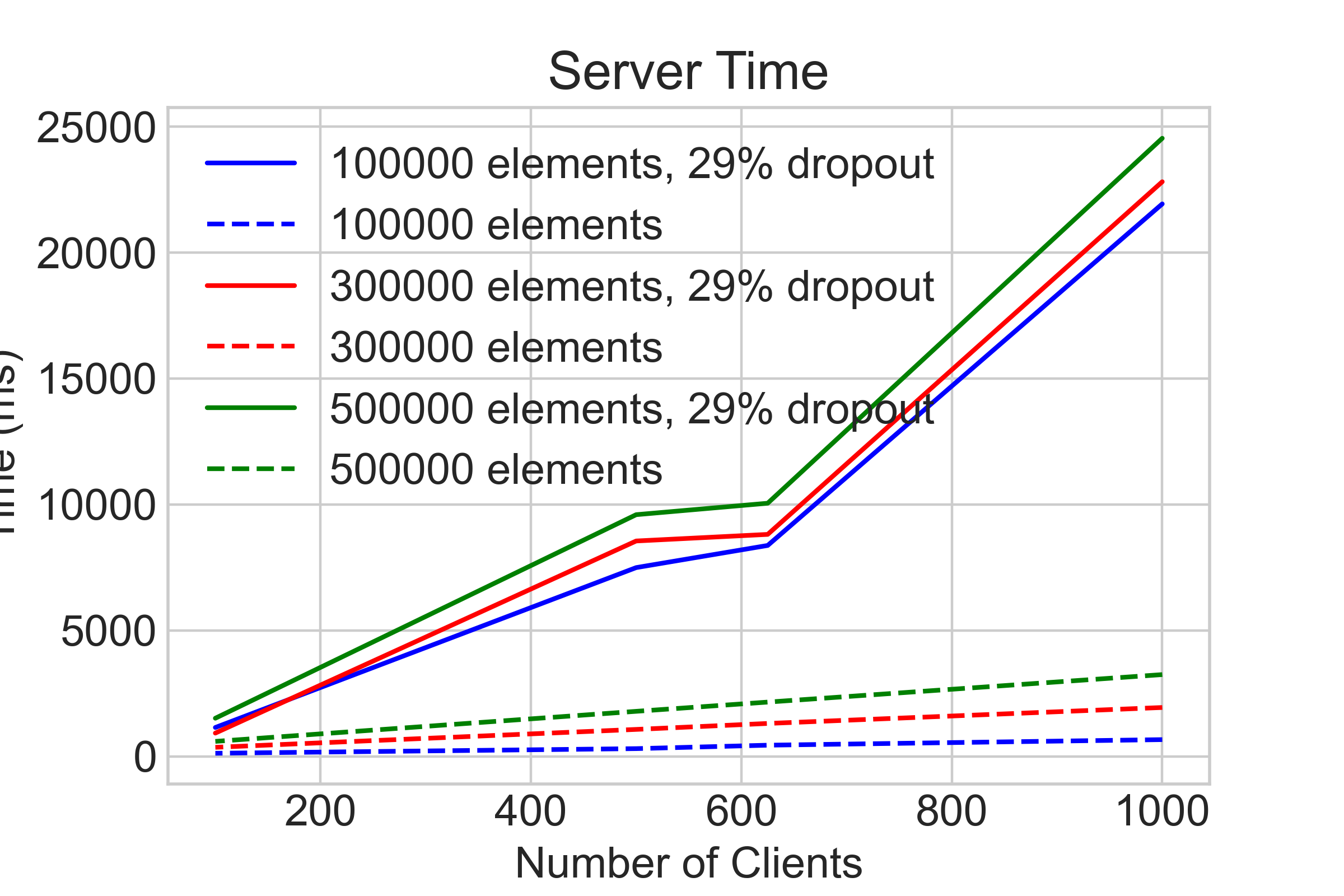}
    \end{subfigure}
    \begin{subfigure}{.5\textwidth}
        \includegraphics[width=.9\linewidth]{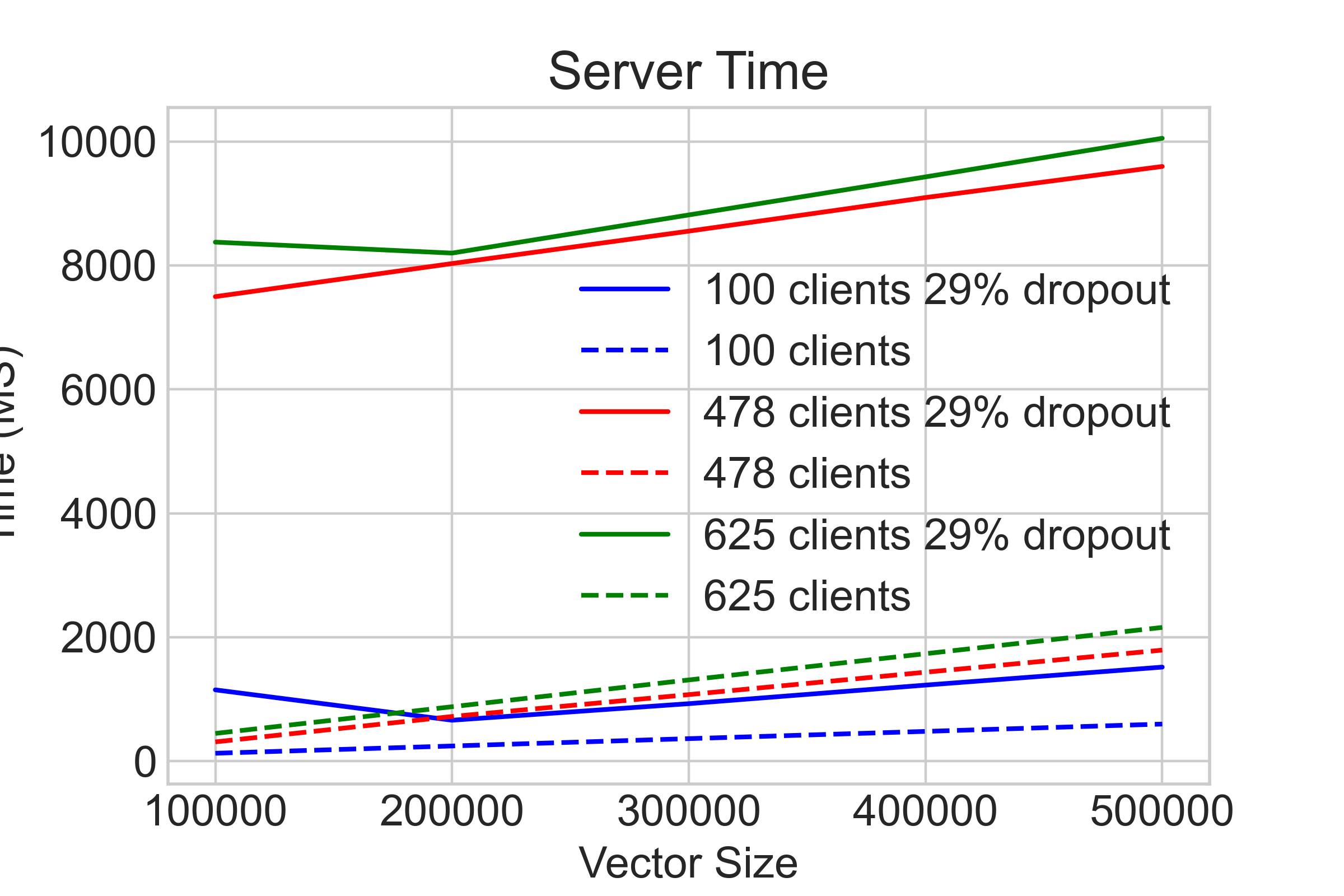}
    \end{subfigure}
    \caption{
    The effects of different federation size and different vector size on server computation time. }
    \label{fig:server:time}
\end{figure*}

Figures~\ref{exp:fig} and \ref{fig:server:time} presents our concrete performance results. We see a significant improvement in client and server computation time over the concrete performance results of Bonawitz et al.~\cite{bonawitz2017aggregation}. Client computation takes less than half a second for all configurations tested, and is dictated by a linear relationship with the vector size. 

Server computation time has a linear relationship with vector size and a quadratic relationship with the number of clients. In the case with no dropouts, server computation is quick, taking less than 5 seconds for all configurations tested. In the dropout scenario, server computation is significantly slower, but still much faster than the state of the art \cite{bonawitz2017aggregation}. It's worth noting that this is an upper bound on server time in the dropout case. Performance can be improved with faster interpolation algorithms \cite{moderncomputeralgebra}.

Recall the quantity $\beta$ from Section \ref{sec:LWEparameters}, which given $q$ the size of the field gives us the standard deviation of the noise. We observe that changing $\beta$ has no effect on the runtime. We note that changing $\beta$ can require different values for $n$ and $q$ to guarantee a certain amount of security, but this is only necessary if $\beta$ is decreased. For our timing experiments we chose $\beta = 3.2/q$ to accommodate a wide variety of privacy budgets for relatively small fixed precision. Because our values are fixed precision with 4 decimal places, the chosen value of $\beta$ adds noise with standard deviation .0409 to our aggregated vectors assuming 128 clients. This is far less than the minimum amount of DP-noise we added in our accuracy experiments, which had a standard deviation of 1. 

\begin{table*}
  \centering
    \def\arraystretch{1.2}
    \begin{tabular}{|c|c|c|c||c|c|c|c|}
    \hline
  Clients & $q$       & $n$  & \% Dropout &  Server &  Client   & Bonawitz Server & Bonawitz Client\\
    \hline \hline
      478 &  31352833 &  710 &   0    &   310   ms &     80   ms &     2018         ms &   849 ms \\
    \hline
      625 &  41057281 &  730 &   0    &   447   ms &     84   ms &     2018         ms &   849  ms \\
    \hline
     1000 &  71663617 &  750 &   0    &   668   ms &     88   ms &     4887         ms &  1699  ms \\
    \hline
    \hline
      478 &  31352833 &  710 &   29   &   496   ms &     91   ms &     143389       ms &   849  ms \\
    \hline
      625 &  41057281 &  730 &   29   &   375   ms &     93   ms &     143389       ms &   849  ms \\
    \hline
     1000 &  71663617 &  750 &   29   &  21931  ms &     99   ms &     413767       ms &   1699 ms \\
    \hline
\end{tabular}

    \caption{Client and server times for various LWE configurations. Vector size is fixed at $100,000$ and $\beta q =3.2$. Times are in milliseconds. Results from Bonawitz et al.~\cite{bonawitz2017aggregation} for 500 and 1000 parties with 0 and 30\% dropout are included for comparison.}
    \label{tab:mask_time}
\end{table*}




\subsection{Experiment 2: Model Accuracy}\label{eval:ml}
\begin{table}
\centering
    \def\arraystretch{1.2}
    \begin{tabular}{|c||c|c|}
        \hline
        \textit{Property} & \textit{MNIST} & \textit{CIFAR-10}\\
        \hline \hline
        Train Set Size & 60,000 & 50,000 \\
        \hline
        Test Set Size & 10,000 & 10,000 \\
        \hline
        \# Conv layers & 2 & 6 \\
        \hline
        \# Parameters & 26,000 & 550,000 \\
        \hline
        Batch Sizes & 16, 32, 64, 128 & 16, 32, 64, 128 \\
        \hline
        $\sigma$ & 0, 1, 2, 4, 8 & 0, 1, 2, 4, 8, 16\\
        \hline
    \end{tabular}
    \caption{Datasets and model configurations used in our experiments}
    \label{tab:modelprops}
\end{table}

This section strives to answer \textbf{RQ2}.
We implement our models in TensorFlow. To preserve privacy, we add noise scaled to a constant $\sigma$ to each example's gradient, which results in the batch gradient described by Equation~\ref{eq:1}. Each gradient is clipped, by a constant $C = 5$ such that batch gradient sensitivity is bounded by $C / \textit{batch\_size}$. These two modifications to a traditional neural network training loop ensure that our models satisfy differential privacy. Adding noise in this way also accurately reflects the process that would be used by a federation member using \ourprot. Gradient updates for individual samples are saved during training for use during the MPC experiments.

We evaluate the accuracy and scalability of \ourprot with the standard MNIST and CIFAR-10 datasets. Both datasets, and the models we train with them are listed in Table~\ref{tab:modelprops}.

For both the MNIST and CIFAR-10 models, we utilize categorical cross entropy for our loss function, stochastic gradient descent with a learning rate of $0.01$ and momentum of $0.9$ for our optimizer and a clipping parameter $C = 5$ for all trials.

\begin{figure*}[!h]
    \includegraphics[width=\textwidth]{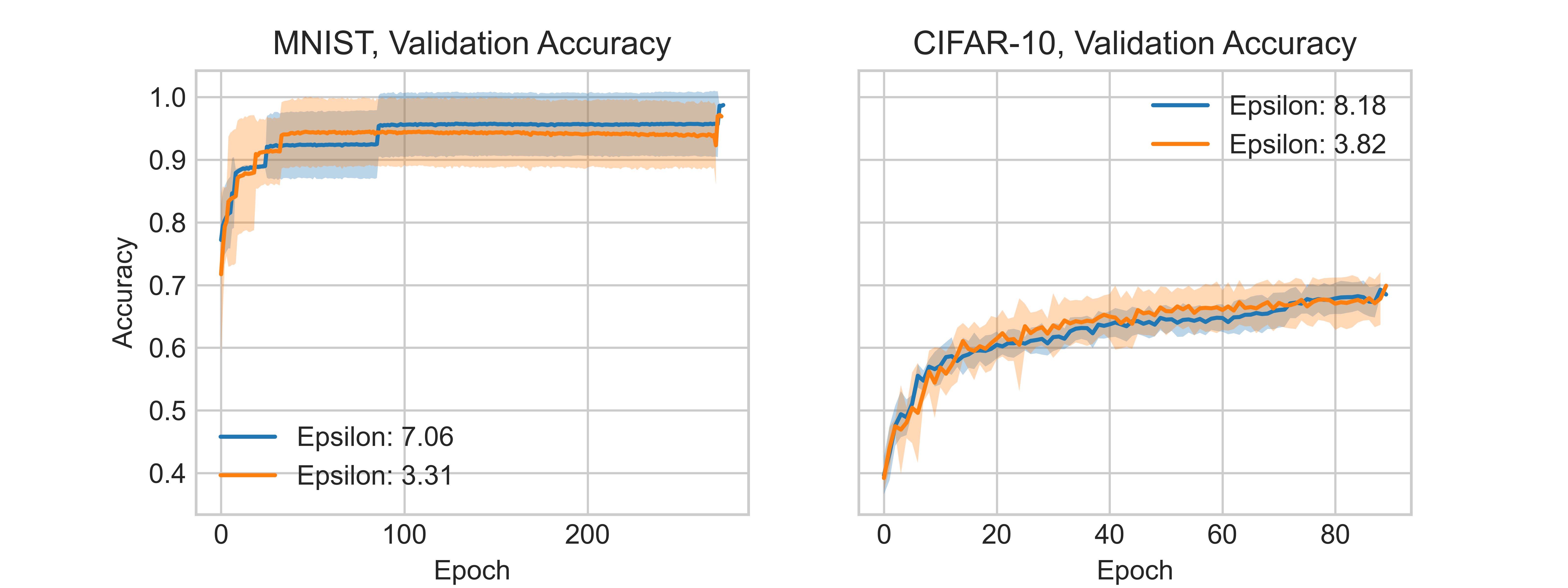}
    \caption{Validation accuracy progression over training runs on MNIST and CIFAR for various values of $\epsilon$ ($\delta = 10^{-5}$). All accuracy values are the average of 4 trials. Batch size is restricted to 64.}
    \label{fig:comb:acc}
\end{figure*}

We run a series of trials for each dataset with each pair of batch size and $\sigma$ listed in Table \ref{tab:modelprops}.
All accuracy results are the per epoch average of 4 trials with the given model configuration.
$\epsilon$ is calculated post hoc as a function of $\sigma, C, \textit{batch\_size}, epochs$.
All $\epsilon$ values are calculated from the corresponding R\'{e}nyi differential privacy guarantee by picking $\alpha$  to minimize the RDP $\epsilon$ parameter, then converting this guarantee into $(\epsilon, \delta)$-differential privacy with $\delta = 10^{-5}$.
We see selected accuracy results reported for differing values of $\epsilon$ in Figure \ref{fig:comb:acc}.

\subsubsection{MNIST}

The Modified National Institute of Standards and Technology database is an often used image recognition benchmark consisting of 60,000 training samples and 10,000 testing samples; each sample is a $28 \times 28$ gray scale image of a handwritten digit. We train a classifier containing 2 ReLU-activated convolution layers, max pooling following each of them, and a ReLU activated dense layer with 32 nodes. Finally, classifications are done with a softmax layer. This model has about 26,000 trainable parameters in total.

\begin{figure*}[!h]
    \begin{subfigure}{.5\textwidth}
        \includegraphics[width=.9\linewidth]{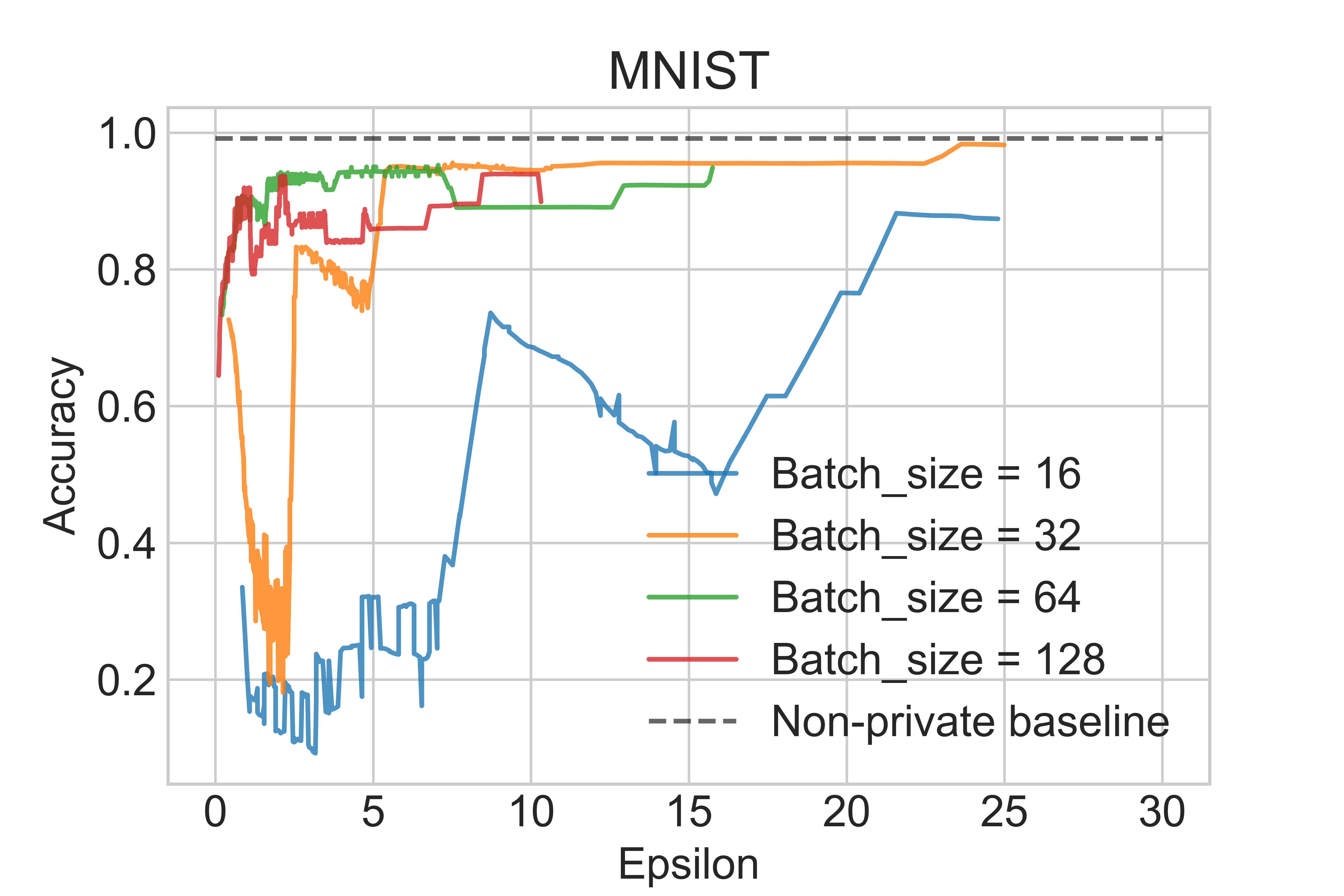}
    \end{subfigure}
    \begin{subfigure}{.5\textwidth}
        \includegraphics[width=.9\linewidth]{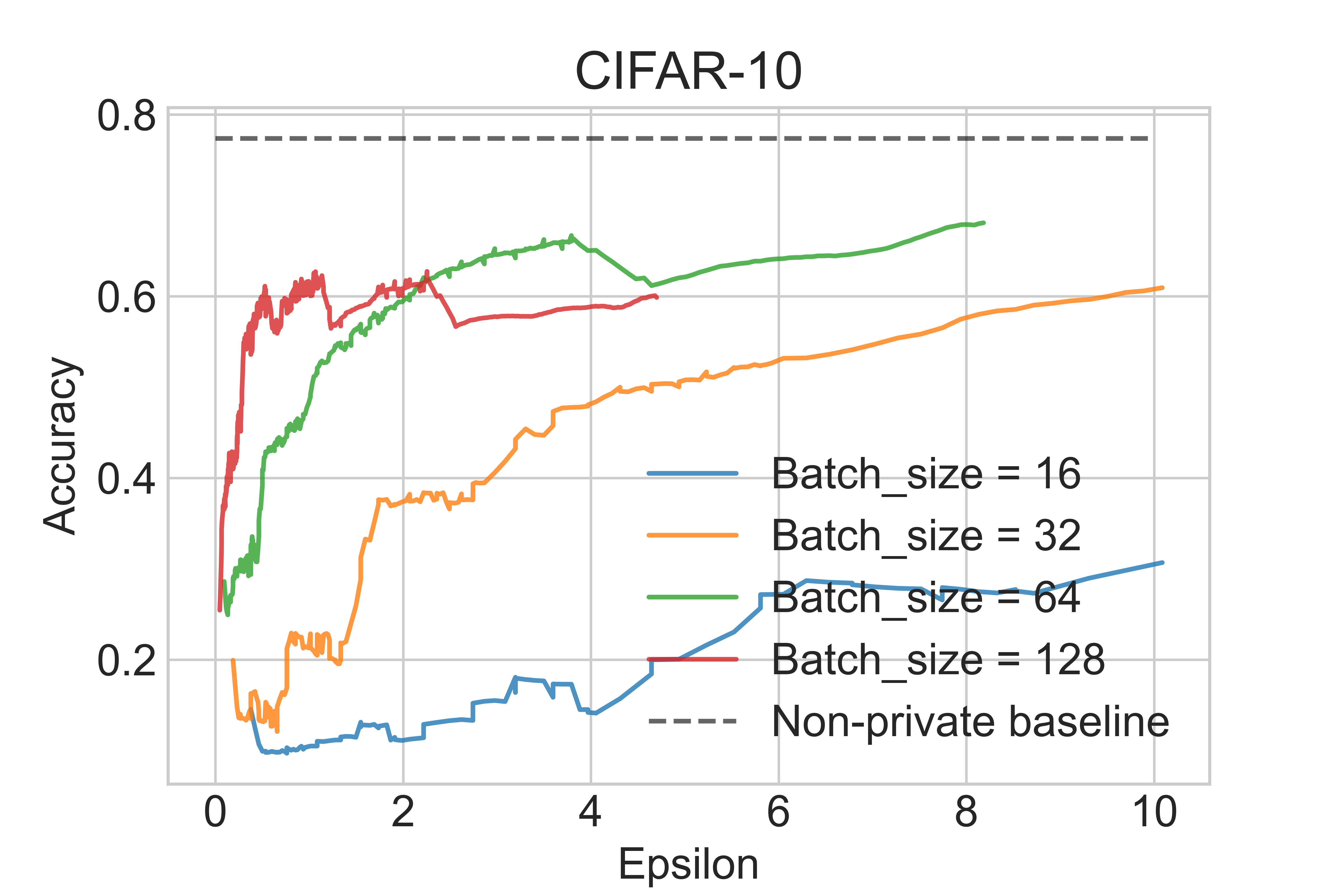}
    \end{subfigure}
    \caption{The effects of privacy budget and batch size on validation accuracy ($\delta=10^{-5}$). Each solid line is a moving average of Accuracy as epsilon increases for a given batch size. The dotted line is the maximum accuracy achieved by our model with no noise added during training. Private federated learning is able to approach non-private accuracy for several batch sizes on MNIST. On CIFAR-10 e see that private models tend to be more accurate with larger batch sizes, while the opposite is true for non-private models.}

    \label{fig:eps-acc}
\end{figure*}

After training for 275 epochs, our private MNIST models are able to attain a maximum $98.7\%$ mean validation accuracy over 4 trials.
This is a slight decrease in accuracy from the no noise baseline accuracy of $99.2\%$, however the private model still generalizes very well.
Figure~\ref{fig:eps-acc} shows how different privacy budgets affect accuracy for our sample batch sizes. 
Models trained with all batch sizes see improved accuracy as $\epsilon$ increases, however larger batch sizes tend to produce more accurate models, especially for small values of $\epsilon$. 
Improved accuracy for larger batch sizes can be seen as an effect of the private average, where the sensitivity of the gradient average is inversely proportional to the batch size. 
Therefore, larger batches require less noise added for a given privacy budget, resulting in a more accurate model.

\subsubsection{CIFAR-10}
 The Canadian Institute for Advanced Research 10 dataset consists of 60,000 colored images equally partitioned into 10 classes. Each image is $32 \times 32$ with 3 channel RGB colored pixels. We separated the dataset into 50,000 training examples and 10,000 test samples for our experiment. Our trained model contains three pairs of ReLU-activated convolution layers with batch normalization after each layer, and max pooling after each pair. We also include one ReLU activated dense layer with 128 nodes, and a softmax activated output layer. This model contains 550,000 parameters.

With a batch size of 64, we achieve a maximum accuracy of $70.0\%$ mean validation accuracy over 4 trials on CIFAR-10.
This is a sizeable drop in accuracy compared to the $77.4\%$ mean accuracy of our architecture trained without differential privacy, however it is in line with differentially private model performance in the central model~\cite{dpdl}. 

Figure~\ref{fig:eps-acc} demonstrates the correlation between larger batch size and greater accuracy when controlling for a specific privacy budget.
As with MNIST, the greater accuracy with larger batch sizes likely stems from gradient sensitivity being dependent on the batch size itself.
That said, for $\epsilon < 10$, we achieve our most accurate model with a batch size of 64 ($\epsilon = 3.67$), which is well within the scalable limits of \ourprot as defined in Section~\ref{eval:mpc}.

\subsubsection{Comparison With Centralized Differential Privacy}

\begin{table}
\centering
    \def\arraystretch{1.2}
\begin{tabular}{|c||c|c|}
    \hline
    \textit{Method} & \textit{Abadi et. al. \cite{dpdl}} & \textbf{\textit{\ourprot}} \\
    \hline \hline
    MNIST ($\epsilon \le 2$) & 95\% & 95\% \\
    \hline
    MNIST ($\epsilon \le 8$) & 97\% & 99\% \\
    \hline
    CIFAR-10 ($\epsilon \le 4$) & 70\% & 70\% \\
    \hline
    CIFAR-10 ($\epsilon \le 8$) & 73\% & 70\% \\
    \hline

\end{tabular}
\caption{A comparison of our private model accuracy with a central-model differentially private training algorithm. For all models, $\delta  = 10^ {-5}$. For all of our models, batch size is $64$.}
\label{tab:dldp}
\end{table}

Our approach produces models with accuracy highly comparable to those achieved by Abadi et. al. \cite{dpdl}. Table~\ref{tab:dldp} shows that for a given privacy budget, our approach is able to produce an output within $3\%$ of the equivalent central-model accuracy. It is worth noting that we report the average of 4 trials in this table, and that we observe the same, or better, decrease in accuracy with respect to the no-noise baseline for each model. These comparable accuracy results demonstrate the usability of \ourprot for privacy preserving federated learning.

%% file: related_work.tex
\section{Related Work}

\paragraph{Secure multiparty computation.}
Secure multiparty computation (MPC)~\cite{evans2017pragmatic} is a family of techniques that enable mutually distrustful parties to collaboratively compute a function of their distributed inputs without revealing those inputs. MPC techniques include \emph{garbled circuits}~\cite{yao1986generate} (which is most easily applied in the two-party case) and approaches based on \emph{secret sharing}~\cite{shamir1979share} (which naturally apply in the $n$-party case). MPC approaches have seen rapid improvement over the past 20 years, but scalability remains a challenge for practical deployments. In particular, most MPC protocols work best when the number of parties is small (e.g., 2 or 3), and costs grow at least quadratically with the number of parties. State-of-the-art protocols support significantly more parties: Wang et al.~\cite{wang2017global} reach 128 parties using a garbled circuits approach, and Chida et al.~\cite{chida2018fast} reach 110 parties using a secret sharing approach.

\paragraph{MPC for differentially private deep learning.}
MPC techniques have been previously applied to the problem of differentially private deep learning, but these approaches require either a {semi-honest data curator}~\cite{truex2019hybrid} or {two non-colluding data curators}~\cite{DBLP:conf/nips/JayaramanW0G18}.
Secure aggregation protocols~\cite{bonawitz2017aggregation, bell_paper} (detailed in Section~\ref{sec:overview}) are themselves MPC protocols, specifically designed for the many-client setting. Kairouz et al.~\cite{kairouz2021distributed} present a general framework for differentially private federated learning that leverages existing secure aggregation protocols.

\paragraph{Security for distributed differential privacy.}
Outside of deep learning, several systems have been proposed for computing differentially private results from distributed data. Honeycrisp~\cite{roth2019honeycrisp} and Orchard~\cite{roth2020orchard} are most related to our work, and use a distributed protocol similar to secure aggregation to compute the results of database-style queries. ShrinkWrap~\cite{bater2018shrinkwrap} and {\sc Crypt$\epsilon$}~\cite{roy2020crypt} leverage existing MPC frameworks to implement differentially private database queries.

\paragraph{Learning with Errors}
As noted in Sections \ref{sec:LWEparameters} and \ref{sec:exp1performance}, in this work we fix $\beta q = 3.2$. We note that the security reductions that ensure that the LWE search problem is difficult do not apply in this case: In \cite{Regev2005}, Regev shows that if $q$ is chosen to be polynomial in $n$, and $\chi$ is a certain discretization of a Gaussian distribution on $\mathbb{F}_q$ with standard deviation $\frac{\beta q}{\sqrt{2\pi}}$ for $0 < \beta <1$ and $\beta q > 2 \sqrt{n}$, then solving the LWE search problem can be quantumly reduced to an algorithm that approximately solves the Shortest Vector Problem and the Shortest Independent Vectors problem. In \cite{Peikert2009}, Peikert shows a classical reduction to the (slightly easier) GapSVP problem. 

While as far as we know there are no security reductions for small fixed $\beta q$, at the same time we do not currently know of an attack that takes advantage of a small constant standard deviation. Accordingly, our choice is similar to the choice made in the current FrodoKEM algorithm specifications (submission to Round 3 of the NIST PQC challenge) \cite{prefrodo, frodo} and consistent with the recommendation of \cite{lwestandards}.

%% file: conclusion.tex
\section{Conclusion}\label{sec:conc}
 In the past decade, an explosion in data collection has led to huge strides forward in machine learning, but the use of sensitive personal data in machine learning also represents a serious privacy concern. 
 We present an approach based on a new protocol called $\ourprot$ that ensures differential privacy for the trained model, \emph{without} the need for a trusted data aggregator.
 Using \ourprot allows a highly accurate model to be trained in a federated (distributed) manner while guaranteeing the privacy of data owners, even against powerful and colluding adversaries. 
 Our empirical results show that these accurate models are trainable within a feasible time frame for practical applications, especially when accuracy and low trust burdens are critical.

The promising results presented in our evaluation also suggest directions for future research.
For example, gradient compression techniques can substantially reduce in-communication overhead for distributed training \cite{DBLPgradcomp}.
Paired with \ourprot, these techniques could further reduce the time per batch for larger models, and potentially improve our scalability with respect to model complexity.
Moreover, we apply \ourprot to the very specific case of privacy preserving federated learning, but
additional research could consider how these techniques scale with simpler, yet important, data problems.
For example, the core noise addition and secure aggregation methods described in this paper could
be adapted to privacy-preserving database queries, while eliminating the need for a central database.

%% file: proof_appendix.tex
\section{Proof of security}\label{app:proof}

Suppose the ideal functionality of noisy vector addition as $F$, an adversary $A$. Let $v_i$ and $x_i$ be input and view of client $i$ respectively. Let $x_s$ be the view of the server. $n$ is the LWE security parameter. Suppose a maliciously secure aggregation protocol $\texttt{Sagg}(X, t)$.  Let $V$ be the output of $\pi$.

Let $U$ be the set of clients, and $C \subset U \cup  \{S\}$ be the set of corrupt parties.

In the malicious model, we consider dropping out an adversarial behavior without loss of generality.

Suppose the simulator has access to an oracle $\texttt{IDEAL}(t, v_u)_{u \in U \setminus C}$ where:

\[ \texttt{IDEAL}(t, v_u)_{u \in U \setminus C} = \begin{cases}
    \Sigma_{u \in U \setminus C} v_u & | U \setminus C | >  t \\
    \bot & otherwise
\end{cases}
\]

Let $\texttt{REAL}_{\pi,C}^U = \{x_i | i \in C\},V$.

\begin{theorem}
     There exists a PPT simulator \texttt{SIM} such that for all $t$, $U$, $C$

\[ \texttt{REAL}_{\pi, C}^U(n, t; v_{U \setminus C}) \equiv \texttt{SIM}_C^{U, \texttt{IDEAL}(t, v_u)}(n, t; x_C) \]
 
\end{theorem}

Proven through the hybrid argument.

\begin{enumerate}
    \item This hybrid is a random variable distributed exactly like $\texttt{REAL}_{\ourprot, C}^U(n, t; v_{U \setminus C})$
    \item In this hybrid $\texttt{SIM}$ has access to $\{x_i | i \in U\}$. $\texttt{SIM}$ runs the full protocol and outputs a view of the adversary from the previous hybrid.
    \item In this hybrid, $\texttt{SIM}$ has corrupt parties receive an $\texttt{ABORT}$ if the server sends a $U_1$ such that $ t > |U_1|$. 
    \item In this hybrid, $\texttt{SIM}$ replaces $V$ with the output of $F$ from any $x_C$.
    \item In this hybrid, $\texttt{SIM}$ generates the ideal inputs of the corrupt parties using the $\texttt{IDEAL}$ oracle, $\texttt{SIM}$ generates a set of random inputs $V_C$ such that $\Sigma_{i \in C} v_i = F(v_U) - \texttt{IDEAL}(t, v_u)_{u \in U \setminus C}$. The output domain of \ourprot is any vector $V \in \mathbb{F}_q^m$ and $ABORT$. $\texttt{SIM}$ can replicate any vector output using this process. Therefore, this hybrid is indistinguishable from the previous hybrid.
    \item In this hybrid, $\texttt{SIM}$ replaces $s$, the sum of secret vectors with a vector of random field elements distributed by $\chi * k$. Because $s$ is not used to reconstruct $G$, and is normally distributed by $\chi *k$, this hybrid is indistinguishable from the previous hybrid. 
    \item In this hybrid, $\texttt{SIM}$ replaces $H$ with $V + As$.
    \item In this hybrid, $\texttt{SIM}$ replaces the run of protocol $\texttt{Sagg}$ with the ideal simulation of $\texttt{Sagg}$.  If $\texttt{Sagg}$ returns $\texttt{ABORT}$, $\texttt{SIM}$ returns $\texttt{ABORT}$. Because $\texttt{Sagg}$ is secure, this hybrid is indistinguishable from the previous hybrid using each parties $s_i$ as input.
    \item In this hybrid, $\texttt{SIM}$ replaces the $s_i$ of each client with a vector of elements distributed by $\chi$. Because $s_i$ is typically distributed by $\chi$ and each $s_i$ is not used to compute $s$ anymore, this hybrid is indistinguishable from the previous hybrid.
    \item In this hybrid, $\texttt{SIM}$ replaces the $b_i$ of each client with a vector of uniformly distributed field elements in $\mathbb{F}_q^m$. Given the LWE assumption, $b_i$ should be indistinguishable from random field elements, so this hybrid is indistinguishable from the previous hybrid from the perspective of the adversary.
    \item In this hybrid, $\texttt{SIM}$ replaces $h_i$ of each client with a vector of uniformly distributed field elements in $\mathbb{F}_q$. By the definition of one time pad, this hybrid should be indistinguishable from the previous hybrid. Additionally this hybrid does not use any input from the honest parties and thus concludes the proof.

\end{enumerate}

After these steps, the simulator no longer needs any input from the honest clients to simulate Protocol~\ref{prot:mask}, implying that it is secure in the malicious threat model.

Notably, our malicious threat model subsumes the semi-honest threat model. Therefore this proof proves security in that threat model as well. In the case of a semi-honest threat model, the security of $\texttt{Sagg}$ can also eased to semi-honest.